\documentclass[11pt]{article}

\usepackage[final]{acl}

\usepackage{times}
\usepackage{latexsym}
\usepackage{enumitem}

\usepackage[T1]{fontenc}
\usepackage[utf8]{inputenc}

\usepackage{microtype}

\usepackage{inconsolata}

\usepackage{graphicx}

\title{Design Techniques for LLM-Powered Interactive Storytelling: \\A Case Study of the Dramamancer System}

\author{Tiffany Wang \quad Yuqian Sun \quad Yi Wang \\ \textbf{Melissa Roemmele \quad John Joon Young Chung \quad Max Kreminski} \\
        Midjourney \\ \{twang, ysun, ywang, mroemmele, jchung, mkreminski\}@midjourney.com}

\begin{document}
\maketitle
% \begin{abstract}
% The rise of Large Language Models (LLMs) has enabled a new paradigm for bridging authorial intent and player agency in interactive narrative. We consider this paradigm through the example of Dramamancer, a system that uses an LLM to transform author-created story schemas into player-driven playthroughs. We outline some design techniques and evaluation considerations arising from this system.
% \end{abstract}

\section{Introduction}

Interactive narrative is a game-like storytelling medium where stories envisioned by authors are realized dynamically through player choices. Traditionally, this medium has posed a challenge to authors in requiring them to anticipate the impact of many possible player choices on the resulting story. The rise of Large Language Models (LLMs) as powerful text generators has suggested a new paradigm for interactive narrative, which imagines that LLMs will eliminate authorial burden by readily instantiating story text that accommodates player agency within narrative boundaries defined by the author \cite[e.g.][]{lu2025,zhu2023}. However, open questions remain about how to best leverage LLMs for this objective. In particular, as LLMs are increasingly noted for their story generation capabilities \cite[e.g.][]{huot2025agents,xie-riedl-2024-creating,yuan2022}, what should the author, player, and LLM each be expected to contribute to the emerging story? To consider this, we look at the example of Dramamancer, a system that utilizes an LLM to transform author-created story schemas into interactive playthroughs driven by player input \cite{sun2025drama,wang2025dramamancer}. We outline some design techniques exemplified by Dramamancer and identify some questions about how to evaluate the stories that emerge from it.

\section{User Interface Design}

The Dramamancer system involves two distinct user roles: an \textbf{author} who creates stories and a \textbf{player} who experiences them. In particular, the author establishes a \textbf{story schema}, and the player engages with an interactive \textbf{playthrough} that dynamically realizes this schema as a concrete story instance based on the player's input. Below, we outline the interface for each of these user roles in terms of what is provided to the system.

\paragraph{Author Interface} A \textbf{story schema} includes:

\begin{itemize}[noitemsep,nolistsep,leftmargin=*]
\item{\textbf{Style}}: The style consists of instructions for the structure and presentation format of the playthrough (e.g. \textit{``Write in Early Modern English with flowery metaphors''}).

\item{\textbf{Characters}}: Each character is defined by a name and description (e.g., \textit{``Sam: a young, novice reporter covering her first superhero competition''}). The first character in the list is designated as the \textbf{player character} represented by the player's input during the playthrough.

\item{\textbf{Scenes}}: Playthroughs are divided into one or more scenes. The author specifies the name of the scene, the characters present (a subset of the above list), a \textbf{setting} establishing the context (e.g. \textit{``competition arena where superheroes face off''}), an \textbf{opening line} that is always presented verbatim to the player at the onset (e.g. \textit{``you walk into the reporter's corner on your first day — the superhero showdown is about to begin, and you want the story.''}), and one or more \textbf{events}.

\item{\textbf{Events}}: Within a particular scene, \textbf{events} define what should happen in the story according to the player's input. Events are structured as \textbf{storylets} \cite{kreminski2018}, each of which consists of a \textbf{condition} and an \textbf{outcome}. A condition is a statement that evaluates to true/false based on the playthrough state (e.g. \textit{``Sam (player) asks what's going on''}). A condition can be null and instead set to trigger after a certain number of lines have been generated in the scene. An outcome is a description of what should happen when the condition is satisfied (e.g. \textit{``veteran reporter explains the competition while heroes argue on stage''}). An outcome can also be configured to end the scene, and additionally transition to a different scene. 
\end{itemize}

\paragraph{Player Interface} A \textbf{playthrough} unfolds line by line. A single \textbf{line} consists of actions and dialogue for one character, which can be accompanied by third-person narration (e.g. ``\textbf{The challenge has begun.} \textit{Maria: (whispering) Which one catches your eye?}'') After certain lines, the playthrough pauses to elicit input from the player. They specify their input for the player character in the format of \textit{(actions) dialogue}. The playthrough for a given scene continues as long as there is no triggered outcome indicating that the scene should end. Once the scene ends, the playthrough proceeds to the next scene if a transition is specified, otherwise the entire playthrough ends.

\section{System Design}

A Dramamancer playthrough is orchestrated by two LLM-based modules. The \textbf{instantiation} module produces the next line of the playthrough, while the \textbf{interpretation} module determines if the current playthrough leads to any of the event conditions being satisfied. Each of these modules is implemented with a single LLM interaction. Below we describe the design features of the prompts for these interactions, which rely on LLMs' general ability to perform tasks in response to instructions and demonstrative examples.

\paragraph{Instantiation} The instantiation prompt facilitates the LLM to generate the next line based on all previously generated lines in the playthrough, conditioned on the style, scene characters, and scene setting. The prompt specifies some of the following expectations for what the LLM will return:

\begin{itemize}[noitemsep,nolistsep,leftmargin=*]
\item The output consists of the line shown to the player as well also other information used to configure the playthrough. This includes a \textit{pause} variable specifying (as a boolean true/false) if the playthrough should pause after that line in order to elicit input from the player. This is expected to be true when there has been meaningful progression in the playthrough since the last time the player contributed input. 
% The line uses notation to distinguish character actions/dialogue from third-person narration.
% \item The line should adhere to the style guidelines.
\item In generating a line, the LLM implicitly decides which character is contributing that line. The line should only pertain to non-player characters, since actions/dialogue for the player character should come only from the player input.
\item Regarding events whose conditions have been satisfied, the prompt lists the outcomes of these events, and instructs the LLM to incorporate them organically into the playthrough. This influences the next line to address outcomes that are not yet reflected in the previously generated lines, while also allowing for the possibility of using multiple lines to convey an outcome.
\item The line should be highly responsive to the most recent input from the player character, regardless of whether the player's input causes any event conditions to be satisfied.
% \item The \textit{plan} variable consists of story-relevant ``notes'' (e.g. character motivations and emotional states) that support cohesion in the subsequently generated lines in the playthrough, but are intentionally hidden from the player.
\end{itemize}

\paragraph{Interpretation} The interpretation module runs after every player input. The prompt facilitates the LLM to determine if a current playthrough satisfies any of the conditions for events in the ongoing scene. The LLM returns a list of satisfied conditions, which enables the outcomes for these events to become available to the instantiation module.

\section{Evaluation Considerations}

The quality of a Dramamancer playthrough is mediated by the performance of the above LLM modules. There are two dimensions to evaluating this quality, corresponding to the perspectives of the author and the player. 

\paragraph{Author Perspective} Author-side evaluation primarily concerns the alignment between the story schema and the playthrough. Some of the specific variables relevant to assessing this are \textbf{style adherence} (does the playthrough consistently follow the style?), \textbf{character distinctiveness} (does the action/dialogue attributed to each non-player character reflect how they are described?), \textbf{scene awareness} (within a given scene, does the playthrough accurately convey the setting?), \textbf{event detection accuracy} (do the scene events get triggered appropriately based on player input?), and \textbf{outcome realization} (are event outcomes expressed suitably based on how they are described?)

\paragraph{Player Perspective} Variables relevant to player-side evaluation include \textbf{responsiveness} (does the player feel like their input is directly influencing the playthrough?), \textbf{timing} (does the playthrough elicit input from the player at the times it is most impactful?), \textbf{reflection}  (does the player feel like they are making meaningful decisions when providing input?), and \textbf{engagement} (does the playthrough motivate the player to stay engaged with the story?).

\section{Summary}

We examine the Dramamancer system as an example of how to utilize LLMs for bridging authorial intent and player agency in interactive narrative. In presenting this case study, we hope to engage further discussion of how evolving LLM capabilities can shape the future of storytelling.


\bibliography{custom}

% \appendix

% \section{Example Appendix}
% \label{sec:appendix}

% This is an appendix.

\end{document}